\documentclass[%
 reprint,
 amsmath,amssymb,
 aps,
 prx,
]{revtex4-2}
\usepackage{graphicx}
\usepackage{dcolumn}
\usepackage{bm}
\usepackage{ulem}
\usepackage{xcolor}
\usepackage{mathrsfs}
\usepackage{soul}

\newcommand{\Lind}{\mathcal{D}}
\newcommand{\brra}[1]{\langle #1|}
\newcommand{\keet}[1]{|#1\rangle }
\newcommand{\sch}{Schrödinger}
\def\SB#1{\color{red}{#1} \color{black}}

\begin{document}

\preprint{APS/123-QED}

\title{Non-Hermitian dynamics in a driven-dissipative and spatially correlated \\ light-matter system.} 

\author{Ruiyang Huang, Eliott Beraud, Antoine Glicenstein, Simon Bernon}
\affiliation{LP2N, Laboratoire Photonique, Numérique et Nanosciences,\\
Université de Bordeaux-IOGS-CNRS: UMR 5298, rue F. Mitterrand, F-33400 Talence, France
}%

\date{\today}

\begin{abstract}
Pure quantum states, as described in quantum mechanics textbooks, are ideal representations inevitably deteriorated in real systems by any dissipative connection to the environment. In this work we derive a unified theoretical frame to study the non-Hermitian dynamics of a driven-dissipative quantum system and validate it by comparison to experimental data of matter-wave diffraction. To highlight the role of dissipation, we perform the comparison in a near-resonant regime of light-matter interaction, where perturbative approaches fail. Here, we show that the developed formalism enables both exact simulations and an intuitive interpretation based on modal analysis. The theoretical analysis is based on the general master equation description of driven-dissipative systems, from which we derive an effective complex potential $V_\mathrm{eff}(\Omega,\Delta)$ that depends on control parameters such as the amplitude and detuning of the light matter interaction. Experimentally, we drive a $^{87}$Rb BEC near resonance using periodic excited-state engineering, mapping the spatially modulated $V_{\rm eff}[\Omega,\Delta(x)]$ into momentum space for precise quantification. The experimental results agree with numerical simulations of the master equation and demonstrate the interplay between coherent drive and dissipation, with an exceptional signature: a reduced decay rate with increasing drive. Such dynamics is then interpreted by a modal analysis of the non-Hermitian Hamiltonian $H_\mathrm{eff}=p^2/2m+V_\mathrm{eff}$ which provide an intuitive and qualitative explanation for such driven-dissipative system.

\end{abstract}

\maketitle


\section{\label{sec:intro}Introduction}
Coherent manipulation of quantum systems dictated by the Hermitian evolution forms the backbone of quantum engineering. Being deterministic, information-preserving and complete, ideal closed-system dynamics provides an optimal framework for quantum computation algorithms. \cite{griffiths_QM,zurek_classical_quantum,DiVincenzo}. In practice, however, perfectly-isolated quantum systems do not exist, making it essential to understand both the consequences and opportunities offered by  open systems. While openness inherently induces information leakage — a feature traditionally minimized — it has recently emerged as a powerful tool. Open systems serve not only as a rich platform to study novel physical phenomena~\cite{NonHPhys_ashida}, including topological protection~\cite{topoclass_general_2025} and critical universality~\cite{Universal_driven_Diehl}, but also enable various techniques such as driven-dissipative state preparation~\cite{Topo_withAtom_zoller} and quantum Zeno state protection~\cite{Mirrahimi_2014}. Driven-dissipative systems, which simultaneously exhibit quantum coherence driven by an external source and dissipation, are of particular interest for probing non-equilibrium quantum physics~\cite{Driven_open_atom_diehl}. Among various platforms, ultracold atoms driven by laser fields offer an exceptionally clean setting for driven-dissipative studies, largely due to the dominant and fully modeled nature of their dissipation channel: spontaneous emission via electromagnetic coupling to the vacuum modes~\cite{NonHPhys_ashida,Universal_driven_Diehl}.


Despite intensive studies, previous investigations of the atomic responses to radiation have largely segregated real and imaginary effects in two distinct regimes. In far-detuned fields, the coherent level shift associated with the state dressing is primarily considered, whereas near resonance, the dynamics are typically approximated by purely dissipative processes. While mechanical effects on thermal atoms have been extensively studied to optimize laser cooling~\cite{DressedApproach_dalibard1985,Dalibard_lectureLaserCool}, the coherent response of ultracold atoms driven near resonance has received comparatively little attention. Although this segregation is operationally practical, the reality is a continuous transition between regimes as a function of detuning. Such continuous treatment has long been adopted to describe the twin classical situation: the effect of an atomic medium on light, accounted for by a complex refractive index $n = n' + i n''$~\cite{Atom_refractive_1995,Refractive_exp2009_single,Refractive_Exp2016_dense,Refractive_Exp2026} which encapsulates both the dispersive (coherent) and extinction (dissipative) responses.


In this work, we bridge this gap by establishing a unified, non-perturbative formalism to describe the complex atomic response especially in the near-resonant driving regime.
Such a unified description is crucial for both fundamental physics and practical quantum technology applications. From a scientific perspective, novel physical regimes emerge in driven-dissipative open systems~\cite{NonHPhys_ashida,topoclass_general_2025,Universal_driven_Diehl}. In ultracold atomic systems, this includes superradiant phase transitions~\cite{superradiantF_2021,superradiance_2023}, dissipative time crystals~\cite{TimeCrystal_2021,TimeCrystal_2}, non-trivial non-Hermitian topological phases in optical lattices~\cite{TopoNonH_Bergholtz,Topo_withAtom_zoller}, and quantum many-body dynamics characterized by correlation-induced stretched-exponential decay of weak excitations~\cite{sketch_expo_2023,Univer_dissi_wenlan,nonH_linear_theory_zhaihui}. Accessing and predicting the behavior of these systems requires a rigorous, quantitative modelling of the atomic system driven close to resonance.

From an engineering perspective, external control channels unavoidably introduce coupling to the environment. Consequently, reaching stronger driving forces pushes the system into the regimes where dissipation cannot be treated perturbatively. In light-matter interactions, non-perturbative dissipation becomes dominant when pushing spatial resolutions beyond the optical diffraction limit or when driving the system with Rabi frequencies comparable to the natural linewidth. Under a "dissipative" interpretation, such sub-wavelength spatial manipulation has been achieved, for example, by excited-state engineering to selectively flip spin with radio-frequency fields~\cite{single_spin_dressing,dressing_rydberg_array} or to spatially modulate a repumping transition achieving excitation resolutions below $100$~nm~\cite{Veyron_2024,Brantut_odt_tomo,Bertoldi_cavity_tomo}. In the "coherent" interpretation, strong electric dipole couplings have been exploited to shape optical traps with sub-wavelength widths~\cite{VaccumForce_Chang,Maxime_2018,bubble}. Neither the dissipative nor the coherent interpretation is fully satisfactory for modeling these regimes that fundamentally require an analytical and numerical framework that extends beyond standard perturbative assumptions.


Here, we present a theoretical and experimental study of the non-Hermitian dynamics of an atomic ensemble subjected to a near-resonant laser field. Based on a Lindblad master equation description of the non-Hermitian evolution, we derive the complex atomic response, which we numerically simulate using the stochastic quantum-jump algorithm (stochastic \sch{} equation) \cite{MCWF_dalibard, MCWF_Zoller}. We verify that the predicted coherent level shifts and dissipation rates consistently satisfy the fundamental constraints of the Kramers-Kronig relations. Experimentally, we probe these internal dynamics by applying a spatially periodic drive that acts as a matter-wave diffraction grating. The observed diffraction patterns confirm the persistence of quantum coherence throughout the driven-dissipative evolution. The loss of contrast as a function of time is a signature of the residual dissipation that our numerical simulations are quantitatively reproducing. Finally, we demonstrate that an eigenmode analysis of the effective non-Hermitian Hamiltonian provides qualitative physical insights into the diffraction dynamics with significantly reduced computational cost compared to full quantum-jump simulations.


This report is organized as follows: In Sec. \ref{sec:th}, we formulate the theoretical model for the effective complex dressing potential experienced by near-resonantly driven atoms, comparing our master equation approach with the traditional mean-force formalism. In Sec.~\ref{sec:exp}, we describe our experimental setup and demonstrate matter-wave diffraction, directly comparing the experimental results with theoretical predictions of both the master equation approach and the mean force formalism. This section includes as well an analysis of the complex eigenmodes of the system which beating and lifetime (encoded in the imaginary part of their  eigen-energies) leads to a comprehensive understanding of the observed temporal dynamics. {Finally, in Sec.~\ref{sec:persp}, we conclude and propose further possible applications of this near-resonant detuning-modulated system as a handy tool in engineering driven-dissipative light-matter interaction.}

\section{Effective Non-Hermitian Hamiltonian based on the Lindblad Master equation}\label{sec:th}

\subsection{Definition of the atomic system}

\begin{figure*}
    \centering
    \includegraphics[width=1\linewidth]{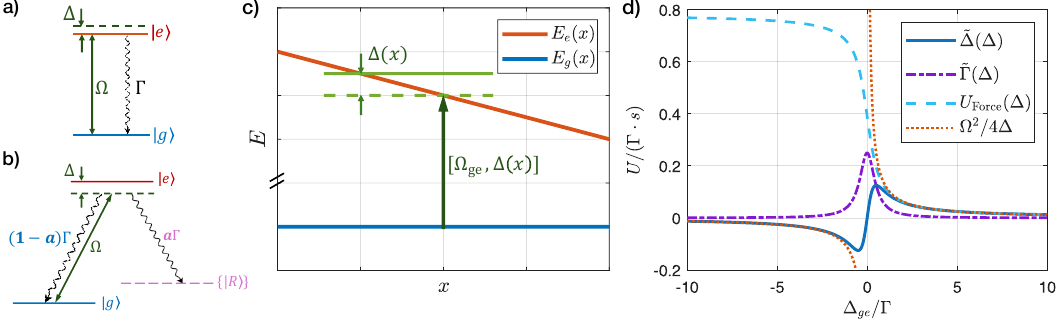}
    \caption{a) Schematic of level structure of 2-level system; b) Schematic of general Raman decay, where part of the atoms decays to a collection of reservoir states; c) Schematic of the excited state engineering with linearly spatial-modulated detuning; d) The dressed ground state potential as a function of detuning, corresponding to Schematic C.}
    \label{fig:level}
\end{figure*}

We initially describe the atoms as an ensemble of closed two-level systems with internal states $\{|g\rangle,~|e\rangle\}$ of respective energies $E_e$ and $E_g$, such that $\omega_0 = (E_e-E_g)/\hbar$ is the atomic transition frequency. This transition is driven by a laser with a detuning $\Delta = \omega - \omega_0$ and a Rabi frequency $\Omega$, as schematically shown in Fig. \ref{fig:level}a). Atoms in the state $|e\rangle$ spontaneously decay towards state $|g\rangle$ occurs at the rate $\Gamma$.

We model the driven-dissipative light-matter interaction using Lindblad maser equation \cite{Davies_1974_master,Lindblad_1976_master}:
\begin{equation}\label{eq:obe}
\begin{aligned}
    \partial_t\rho &= -\frac{i}{\hbar}[H_0+H_\mathrm{int},\rho]+\Gamma\Lind[\sigma]\rho
    \\&= -\frac{i}{\hbar}[H_0+H_\mathrm{int},\rho]-\frac{\Gamma}{2}\{\sigma^\dagger\sigma,\rho\}+\Gamma\sigma\rho\sigma^\dagger
\end{aligned}
\end{equation}
where $\sigma=\keet{e}\brra{g}$, $\Lind[\sigma]$ is the Lindblad superoperator and the free evolution and interaction Hermitian Hamiltonians in the rotating frame are respectively defined as:
\begin{alignat}{2}
    &H_0&&=-\hbar\Delta \sigma^\dagger\sigma ,\label{eq:H0_internal}\\
    &H_\mathrm{int}&&= \Omega\sigma^\dagger+\Omega^*\sigma
\end{alignat}
The atomic dynamics including dissipation is then determined by the parameters $\Delta$ and $\Omega$.

\subsection{The mean force formalism}

To control the external degrees of freedom of the atoms via atom-light interaction, one typically creates a spatial variation of the parameters $\Delta(\mathbf{r})$ and $ \Omega(\mathbf{r})$. A classical example is the optical dipole potential:

\begin{equation}\label{eq:U_dress_fardet}
U(\mathbf{r})=\hbar|\Omega(\mathbf{r})|^2/4\Delta
\end{equation}
which is experimentally realized, in the far-detuned regime $\Delta\gg\Gamma$, by spatially modulating the laser intensity $|\Omega(\mathbf{r})|^2$. 


In earlier works \cite{DressedApproach_dalibard1985}, the near-resonant regime $\Delta\sim\Gamma$ was treated within the framework of the localized mean force, wherein the internal state of a slowly moving atom adiabatically follows the local equilibrium determined by  $\Delta(\mathbf{r})$ and $ \Omega(\mathbf{r})$. Averaging over this local internal equilibrium yields the mean force:

\begin{equation}\label{eq:meanF}
    F = \mathrm{Tr(}\rho\mathbf{F}) = -(\nabla U_e\rho_{ee}^\mathrm{st}+\nabla U_g\rho_{gg}^\mathrm{st})
\end{equation}
where $\rho^\mathrm{st}$ denotes the steady-state solution of the Optical Bloch Equations (Eq. \ref{eq:obe}). This mean force formalism has been thoroughly discussed in \cite{DressedApproach_dalibard1985} and proven successful in the context of laser cooling and trapping of atomic gases. However, the mean force formalism assumes $r =\langle\hat r\rangle$, i.e., that the atoms are thermal particles with well-defined positions. Thus it cannot describe a spatially-coherent matter wave, whose extension in space can be larger than the laser wavelength. 

To extend this formalism to quantum-degenerate systems, \cite{VaccumForce_Chang} proposed approximating the potential by integrating the local mean force. The derivation was formulated in 1D for a system with a spatially modulated detuning $\Delta(x)$ and a uniform intensity $|\Omega|^2$ (Fig. \ref{fig:level}c), yielding:

\begin{equation}\label{eq:Ux_by_int}
    U(x)=-\int F(x) dx.
\end{equation}
Here, $\Delta(x)$ governs the force through the steady-state population matrix elements $\rho^\mathrm{st}$, obtained by setting $\partial_t\rho=0$ in Eq. \ref{eq:obe}.
Applying the chain rule of differentiation to Eq. \ref{eq:meanF} rewrites the force as:

\begin{equation}
    F = -\left(
    \frac{\partial U_e}{\partial\Delta}
    \rho_{ee}^\mathrm{st}
    +\frac{\partial U_e}{\partial\Delta}
    \rho_{gg}^\mathrm{st}\right)
    \cdot\frac{\partial\Delta}{\partial x} = F(\Delta)\cdot\frac{\partial\Delta}{\partial x}
\end{equation}
Substituting this form into Eq.~(\ref{eq:Ux_by_int}) yields:

\begin{equation}
    U(x) = -\int F(\Delta)\cdot\frac{\partial\Delta}{\partial x}dx = -\int F(\Delta)d\Delta.
\end{equation}
Consequently, the potential can be expressed analytically as a function of the local detuning and Rabi frequency as : 

\begin{equation}\label{eq:U_force}
    U_F(\Delta,\Omega) = -\frac{\hbar\Gamma s}{4\sqrt{1+s}}
    \arctan\left(
    \frac{2\Delta/\Gamma}{\sqrt{1+s}}
    \right)
\end{equation}
where $s=2|\Omega|^2/\Gamma^2$ is the saturation parameter at resonance. 
This potential is plotted as a function of $\Delta$ for a fixed $\Omega$ in Fig. \ref{fig:level}d (light blue), where we set the potential offset to match the asymptotic perturbative results in the limit $\Delta \rightarrow +\infty$. A pronounced discrepancy with the perturbative result, $\hbar\vert{}\Omega\vert{}^2/(4\Delta)$, thus emerges in the red-detuned regime ($\Delta < 0$).  
This unphysical asymmetry signals a breakdown of the conservative assumption. Specifically, integrating the mean force around a closed spatial loop yields non-zero work, $\oint \mathbf{F}\cdot d\mathbf{r} \neq 0$, thereby violating energy conservation.

By averaging over the internal dynamics in Eq. \ref{eq:meanF}, the contributions to the mean force from the ground and excited states are considered on equal footing, whereas the excited population $\rho_{ee}$ also undergoes dissipative dynamics. In other words, the derived mean force contains radiation friction components that should not be included in the integration. Therefore, the mean force $F(x)$ cannot be used to derive a conservative potential. Instead, it should only be applied to thermal atoms with well-defined positions $x$. 

\subsection{Master equation formalism}

\subsubsection{Effective Non-Hermitian Hamiltonian for closed two-level atoms}

The goal of this section is twofold: first, we present a master-equation description of the non-Hermitian dynamics of near-resonantly driven atoms. Second, we derive an effective complex potential that offers a unified picture for both the coherent and dissipative light-matter interaction. These two objectives are achieved simultaneously by a mathematical transformation of the master equation (Eq.~\ref{eq:obe}) wherein the commutator and anti-commutator are combined via the definition of an effective non-Hermitian Hamiltonian $H_\mathrm{eff}$:
\begin{equation}
    [H_\mathrm{eff},\rho] = H_\mathrm{eff}\rho-\rho H_\mathrm{eff}^\dagger
\end{equation}
Specifically for Eq. \ref{eq:obe}, this effective Hamiltonian takes the form:
\begin{equation}
    H_\mathrm{eff}:=H_0+H_\mathrm{int}-i\frac{\hbar\Gamma}{2}\sigma^\dagger\sigma
\end{equation}
which simplifies Eq. \ref{eq:obe} into:

\begin{equation}\label{eq:obe_nonH}
    \partial_t\rho = -\frac{i}{\hbar}[H_\mathrm{eff},\rho]+\Gamma\sigma\rho\sigma^\dagger.
\end{equation}
Compared to the standard Lindblad superoperator form  $\Lind[\sigma]$ (Eq. \ref{eq:obe}), Eq.~\ref{eq:obe_nonH} separates the effective Hamiltonian $H_\mathrm{eff}$, which acts directly on state vectors, from the dephasing term $\Gamma\sigma\rho\sigma^\dagger$, which accounts for quantum jumps and cannot be absorbed into a generalized commutator. For now, Eq. \ref{eq:obe_nonH} describes only the internal dynamics. To proceed, we must incorporate the role of external motion into $H_\mathrm{eff}$, from which an effective, spatially varying complex potential $V_\mathrm{eff}$ can be extracted. 

External dynamics of the density matrix $\hat\rho$ are taken into account by including the kinetic term and the explicit spatial dependence of the parameters in its evolution. For simplicity, we restrict the derivation to a 1D formalism, expressed as $x$. These extra terms are thus defined as follows:

\begin{itemize}
    \item The kinetic energy is included in the Hamiltonian $H_0$ as follows:
    \begin{equation}
        H_0=\frac{\mathbf{p^2}}{2m}-\hbar\Delta \sigma^\dagger\sigma
    \end{equation}
    \item The parameters $\Delta,~\Omega$ become spatial operators $\hat\Delta,~\hat\Omega$. Taking the detuning $\Delta(x)$ as an example, its operator form writes:
    \begin{equation}
        \hat\Delta = \int \Delta(x)\keet{x}\brra{x}dx
    \end{equation}
    \item The full density matrix results from the tensor product between internal and external degrees of freedom:
    \begin{equation}
        \hat{\rho} = \sum_{\alpha\beta = g,e}\iint dxdx'\rho_{\alpha\beta}(x,x')\keet{\alpha}\brra{\beta}\otimes\keet{x}\brra{x'}
    \end{equation}
    Accordingly, the spatial density matrix associated with each internal block is defined as:
\begin{equation}
    \hat{\rho}_{\alpha\beta} = \iint dx \, dx' \, \rho_{\alpha\beta}(x,x') |x\rangle\langle x'|.
\end{equation}

\end{itemize}

Spatial modulation of the parameters explicitly breaks the translational invariance. Thus, the commutation relations between the density matrix $\hat\rho$ and the spatially dependent operators ($\hat{\Delta},\hat{\Omega}$) are no longer taken for granted: If spatial correlation is present in the system (the off-diagonal terms $\hat\rho_{\alpha\beta}(x,x')$ is non-zero) and the parameter has spatial variation (take $\Delta(x)\neq \rm{const}$ as example), then $[\hat\rho,\hat\Delta]\neq0$. Substituting these operator definitions into the master equation Eq. \ref{eq:obe}, we obtain the equations of motion for each internal density-matrix component $\hat{\rho}_{\alpha\beta}$:
{\small 
\begin{equation}\label{eq:obeSpace}
    \begin{aligned}
    &\partial_t\hat\rho_{gg} = -\frac{i}{\hbar}\left[\frac{p^2}{2m},\hat\rho_{gg}\right] 
    - \frac{i}{2}\left(\hat\Omega^*\hat\rho_{eg}-\hat\rho_{ge}\hat\Omega\right)+
    \Gamma\hat\rho_{ee} \\
    &\partial_t\hat\rho_{ee} = -\frac{i}{\hbar}\left[\frac{p^2}{2m}-\hbar\hat\Delta,\hat\rho_{ee}\right] 
    + \frac{i}{2}\left(\hat\Omega^*\hat\rho_{eg}-\hat\rho_{ge}\hat\Omega\right)- 
    \Gamma \hat\rho_{ee} \\
    &\partial_t\hat\rho_{ge} = -\frac{i}{\hbar}\left[\frac{p^2}{2m},\hat\rho_{ge}\right] 
    - \frac{i}{2}\left(
    \hat\Omega^*\hat\rho_{ee}
    -\hat\rho_{gg}\hat\Omega^*
    \right)
    - \hat\rho_{ge}\left(\frac{\Gamma}{2}+i\hat\Delta\right)\\
    &\partial_t\hat\rho_{eg} = -\frac{i}{\hbar}\left[\frac{p^2}{2m},\hat\rho_{eg}\right] 
    - \frac{i}{2}\left(
    \hat\Omega\hat\rho_{gg}-
    \hat\rho_{ee}\hat\Omega\right)
    - \left(\frac{\Gamma}{2}-i\hat\Delta\right)\hat\rho_{eg}
\end{aligned} 
\end{equation}}
where strict operator ordering between the spatial density matrices $\hat{\rho}_{\alpha\beta}$ and the spatial operators $(\hat{\Delta}, \hat{\Omega})$ has been preserved, as a direct result of the spatial correlations present in a quantum-degenerate atomic ensemble. This is an important feature of Eqs. \ref{eq:obeSpace} that depicts the time evolution of \textit{both} the internal (e.g. OBE) and external dependences of the density matrix.

The decoherence terms proportional to $\Gamma$ fundamentally distinguish the ground-state component $\hat\rho_{gg}$ from the others. In the latter three equations, $\mathbf{Re}(\partial_t \hat\rho_{\alpha\beta}) \propto -\Gamma\hat\rho_{\alpha\beta}$ causes the amplitudes of $\hat{\rho}_{ee}$, $\hat{\rho}_{eg}$, and $\hat{\rho}_{ge}$ to decay exponentially on the timescale of $\Gamma^{-1}$.
Thus, no phase correlation can be preserved beyond this timescale. 
Typically for alkali atoms, the natural linewidth $\Gamma\sim2\pi\cdot$MHz, yields a decay time $\Gamma^{-1} \lesssim 100\text{ ns}$, which is orders of magnitude faster than the characteristic timescale of external motion ($>$µs). In contrast, the ground state population $\hat\rho_{gg}$ has a coherence which is only affected by the quantum jumps from $\keet{e}$ to $\keet{g}$ that occur at a rate proportional to $\Gamma\hat\rho_{ee}$ (and not to $\Gamma\hat\rho_{gg}$). In the weak-pumping regime ($\Omega^2\ll\Gamma^2$, $\Delta^2\gg\Gamma^2$ or both), $\hat\rho_{ee}\ll\hat\rho_{gg}\approx1$, allowing the ground-state spatial coherence to survive over a timescale comparable to that of the external dynamics. This asymmetry between ground- and excited-state dynamics highlights a key distinction between the master equation approach and the localized mean force treatment. 
Furthermore, the separation of decoherence timescales justifies an adiabatic approximation. To describe the coherent evolution of $\hat{\rho}_{gg}$ over microsecond timescales, we can assume that the internal coherences rapidly relax to their steady states $\hat{\rho}_{eg,ge}^\mathrm{st}$:


\begin{equation}\label{eq:rho_egge^st}
    \begin{aligned}
    &\hat\rho_{ge}^\mathrm{st}=
    - \frac{i}{2}\left(\hat\Omega^*\hat\rho_{ee}-\hat\rho_{gg}\hat\Omega^*\right)
    \left(\frac{\Gamma}{2}+i\hat\Delta\right)^{-1}\\
    &\hat\rho_{eg}^\mathrm{st}= 
    - \frac{i}{2}\left(\frac{\Gamma}{2}-i\hat\Delta\right)^{-1}\left(\hat\Omega\hat\rho_{gg}-\hat\rho_{ee}\hat\Omega\right).
    \end{aligned}
\end{equation}
By substituting these steady-state values into the equation for $\hat{\rho}_{gg}$ in Eq.~\ref{eq:obeSpace} yields:

\begin{widetext}
\begin{alignat}{2}
    \partial_t\hat\rho_{gg}&=-\frac{i}{\hbar}\left[\frac{p^2}{2m},\hat\rho_{gg}\right] 
    -\frac{i}{\hbar}&&\left[ 
    \frac{\hbar|\hat\Omega|^2(\hat\Delta-i\Gamma/2)}{\Gamma^2+4\hat\Delta^2}\hat\rho_{gg} - \hat\rho_{gg}\frac{\hbar|\hat\Omega|^2(\hat\Delta+i\Gamma/2)}{\Gamma^2+4\hat\Delta^2}
    \right. \nonumber \\
    &&&
    \left. 
    -\frac{\hbar(\hat\Delta-i\Gamma/2)}{\Gamma^2+4\hat\Delta^2}\hat\Omega^*\hat\rho_{ee}\hat\Omega
    + \hat\Omega^*\hat\rho_{ee}\hat\Omega
    \frac{\hbar(\hat\Delta+i\Gamma/2)}{\Gamma^2+4\hat\Delta^2}
    \right]
    +\Gamma\hat\rho_{ee}. \label{eq:pt_rhogg}
\end{alignat}
\end{widetext}
The terms proportional to $\hat{\rho}_{gg}$ can be absorbed into the commutator as a first-order complex effective potential $V_\mathrm{eff}^{(1)}$ that neglects the contributions proportional to $\hat\rho_{ee}$ terms:

\begin{equation}\label{eq:veff_1st}
    V^{(1)}_\mathrm{eff} = 
    \frac{\hbar|\Omega|^2(\Delta-i\Gamma/2)}{\Gamma^2+4\Delta^2} = \tilde\Delta - i \tilde\Gamma
\end{equation}
where the real and imaginary parts of $V^{(1)}_\mathrm{eff}$ define the coherent energy shift $\tilde\Delta$ and the effective dissipation $\tilde\Gamma$, respectively.

This effective potential $V^{(1)}_\mathrm{eff}$, which applies to the evolution of the ground state population $\hat\rho_{gg}$, is solely determined by the local parameters $(\Delta(x),\Omega(x))$, even though its derivation relies on the non-commutativity (spatial correlations) between  $\hat\rho_{gg}$ and the operators $\hat\Delta,~\hat\Omega$. Following this principle, we can further incorporate higher order corrections arising from the $\hat\rho_{ee}$ terms in Eq. \ref{eq:pt_rhogg}:

\begin{align}
    \partial_t\hat\rho_{gg}=&-\frac{i}{\hbar}\left[\frac{p^2}{2m},\hat\rho_{gg}\right] 
    \nonumber \\
    &-\frac{i}{\hbar}\left[ 
    V_\mathrm{eff}^{(1)}(\hat\rho_{gg}-\hat\rho_{ee})
    - (\hat\rho_{gg}-\hat\rho_{ee})V_\mathrm{eff}^{(1)}\right]
    +\Gamma\hat\rho_{ee}\nonumber \\
    \approx &-\frac{i}{\hbar}\left[\frac{p^2}{2m}+V_\mathrm{eff}^{(1)}\cdot\frac{\rho_{gg}^\mathrm{st}-\rho_{ee}^\mathrm{st}}{\rho_{gg}^\mathrm{st}},\hat\rho_{gg}\right] +\Gamma\hat\rho_{ee} \label{eq:rho_t_gg_2ls}
\end{align}
which is valid under two approximations:

\begin{itemize}
    \item Assuming $[\hat\rho_{ee},\hat\Omega]\approx 0$, which is exact for uniform laser intensity and remains valid if $\hat{\rho}_{ee}$  loses its spatial coherence (on the timescale $\Gamma^{-1}$) faster than external motion ($\sim$µs), as is the case for alkali transitions; 
    \item Assuming $V_\mathrm{eff}$ is defined locally, so that only \textit{local} steady-state values $\rho_{gg}^\mathrm{st}$ and $\rho_{ee}^\mathrm{st}$ contribute to the higher-order correction, thus giving rise to a scalar correction to the potential $V_\mathrm{eff}$. 
\end{itemize}

The local steady-state population $\rho_{gg,ee}^\mathrm{st}$ is obtained from the steady-state conditions $\partial_t\rho_{ee}=0$ and $\rho_{gg}^\mathrm{st}+\rho_{ee}^\mathrm{st}=1$ :
\begin{equation}\label{eq:rho_ee_st}
    \rho_{ee}^\mathrm{st}=\frac{|\Omega|^2}{\Gamma^2+4\Delta^2+2|\Omega|^2}
\end{equation}
which matches the standard steady-state solution of the Optical Bloch Equations (Eq. \ref{eq:obe}). Substituting Eq. \ref{eq:rho_ee_st} into Eq. \ref{eq:rho_t_gg_2ls} yields the effective potential $V_\mathrm{eff}$ beyond the first order:

\begin{equation}
    V_\mathrm{eff} = 
    \frac{\hbar|\Omega|^2(\Delta-i\Gamma/2)}{\Gamma^2+4\Delta^2+|\Omega|^2}
\end{equation}
where the additional $|\Omega|^2$ term in the denominator accounts for optical saturation.

To summarize, this master-equation framework provides a unified non-Hermitian description of both  conservative and dissipative responses as functions of the local parameters $(\Delta, \Omega)$. 
The real part $\tilde{\Delta}(\Delta)$ and imaginary part $\tilde{\Gamma}(\Delta)$ of the first-order potential $V_\mathrm{eff}^{(1)}$ are plotted in Fig.~\ref{fig:level}d as a function of the detuning $\Delta$ for a given Rabi frequency $\Omega$.
The real part notably recovers the perturbative limit at both large positive and negative detunings. Moreover, the line shapes observed in Fig.~\ref{fig:level}d mirror the dispersive and absorptive profiles experienced by light propagating in an atomic medium. Such effects are modeled by a complex refractive index whose components are related by the Kramers-Kronig relations. Such relations, which apply to the frequency response of any linear system, are also valid and are consistently retrieved from our master equation approach.

The derivation relies on the non-commutativity between the spatial density-matrix components $\hat\rho_{\alpha\beta}$ and the parameter operators $\hat \Delta,~\hat\Omega$. 
This non-commutativity originates directly from the off-diagonal elements of $\hat{\rho}_{\alpha\beta}$, which encode the spatial correlations within the matter wave. This marks a key distinction from the local mean-force formalism, which treats atoms as point-like localized particles.

\subsubsection{Extension to open level structure with Raman decay}

In the preceding section, we derived the analytical equation of motion for the ground-state population $\hat{\rho}_{gg}$ in a closed two-level system.
We now extend this formalism to the more general case of an open transition (Fig.~\ref{fig:level}b), where excited atoms can decay with a branching ratio $a$ into a manifold of reservoir states $\{|R\rangle\}$ which are distinct from the initial ground state $|g\rangle$. The time evolution of $\hat\rho_{gg}$ is thus modified as:
\begin{equation}
    \partial_t\hat\rho_{gg} = -\frac{i}{\hbar}\left[\frac{p^2}{2m},\hat\rho_{gg}\right] 
    - \frac{i}{2}\left(\hat\Omega^*\hat\rho_{eg}-\hat\rho_{ge}\hat\Omega\right)+
    (1-a)\Gamma\hat\rho_{ee}
\end{equation}
reflecting that only a fraction $(1-a)$ of the spontaneously decayed atoms returns to the ground state $\keet{g}$. Following the approach established for the closed two-level system, i.e. separating non-Hermitian Hamiltonian from the dissipation term, we seek to recast the equation in the form: 

\begin{equation}\label{eq:rho_t_gg_R}
    \partial_t\hat\rho_{gg} = -\frac{i}{\hbar}\left[\frac{p^2}{2m}+V_\mathrm{eff}^R,\hat\rho_{gg}\right] 
    +
    (1-a)\Gamma\hat\rho_{ee}
\end{equation}
where $V_\mathrm{eff}^R$ (with superscript $R$) denotes the effective potential corrected for Raman scattering into the reservoir. 

In the case of a closed transition, higher-order corrections were introduced in Eq.~\ref{eq:rho_t_gg_2ls} through the local steady-state populations. An open transition modifies these local steady-state values $\rho^\mathrm{st}$ due to the loss of local atom number:
\begin{equation}
\partial_t(\rho_{gg}+\rho_{ee}) = -a\Gamma\rho_{ee},
\end{equation}
which subsequently leads to a modified higher-order correction for $V_\mathrm{eff}^R$. To determine the updated steady-state values, we define a local loss rate $\kappa = a\Gamma\rho_{ee}^\mathrm{st}$ and adopt a separability approximation $\rho = \rho^Re^{-\kappa t}$, wherein we isolate the internal two-level dynamics from the exponential decay that corresponds to the population loss into the reservoir:
\begin{equation}
\partial_t(\rho^R_{gg}+\rho^R_{ee}) = -a\Gamma\rho_{ee}+\kappa(\rho^R_{gg}+\rho^R_{ee}) = 0,
\end{equation}
with the initial condition $(\rho^R_{gg}+\rho^R_{ee})|_{t=0} =1$. Substituting these updated steady-state components $\rho_{\alpha\beta}^\mathrm{st}$ into the equation of motion for $\hat\rho^R_{gg}$ (Appendix~\ref{app:raman}) yields the corrected effective potential:
\begin{equation}\label{eq:veff_r}
V_\mathrm{eff}^R = \frac{\hbar|\Omega|^2[\Delta-i(\Gamma-2\kappa)/2]}{(\Gamma-2\kappa)^2+4\Delta^2}\left(1-\frac{\rho_{ee}^\mathrm{st}}{\rho_{gg}^{st}}\right),
\end{equation}
along with a self-consistent set of equations for $\kappa$ and $\rho_{ee}^{st}$:
\begin{equation}
\begin{aligned}
\kappa &= a\Gamma\rho_{ee}^\mathrm{st}, \\
\rho_{ee}^\mathrm{st} &= \frac{1}{2}\frac{s^R}{\left(1+\frac{\kappa}{\Gamma-2\kappa}\right)+s^R}, \\
s^R &= \frac{2|\Omega|^2}{(\Gamma-2\kappa)^2+4\Delta^2}.
\end{aligned}
\end{equation}
Using this self-consistent system, $V_\mathrm{eff}^R$ can be solved numerically for any branching ratio $a$. Beyond these higher-order modifications to $V_\mathrm{eff}^R$, a key physical distinction lies in the non-conservation of atom number within the $\{\keet{g}, \keet{e}\}$ subspace (Eq.~\ref{eq:rho_t_gg_R}). Dissipative events induce not only decoherence, but also atom loss from this two-level subspace — a quantity easier to measure experimentally than spatial coherence. We therefore used an open transition in our experiments to track dissipation events.

\subsubsection{Numerical Simulation Methods}
As mentioned above, the non-Hermitian Hamiltonian can be directly applied to a state vector. When the generalized commutator $[H_\mathrm{eff},\rho]$ is the only term in the master equation — such as in pure Raman decay with $a=1$ in Eq.~\ref{eq:rho_t_gg_R} — a Schrödinger equation with non-Hermitian Hamiltonian fully describes the system. This reduction from a density matrix to a state vector significantly reduces computational complexity. To include dissipation in a state vector formalism, \cite{MCWF_dalibard, MCWF_Zoller} have developped the Stochastic Schrödinger approach (Monte Carlo Wave Function) where dissipative events are treated as discrete, stochastic quantum jumps. By averaging over independent quantum trajectories, this method recovers the full master-equation dynamics. Although stochastic for individual trajectories, it is computationally efficient for large systems and converges rigorously to the master-equation solution as the number of repetition increases. We apply this method to simulate our experiments and compare the results with measured diffraction dynamics in Fig.~\ref{fig:time_evo}. Details on the computational method are given in Appendix~\ref{app:numerical}.

\section{Experimental Realization in a Detuning-Modulated System}\label{sec:exp}
\subsection{Experimental Design}\label{sec:exp_setup}
\begin{figure*}
\centering
\includegraphics[width = 0.99\linewidth]{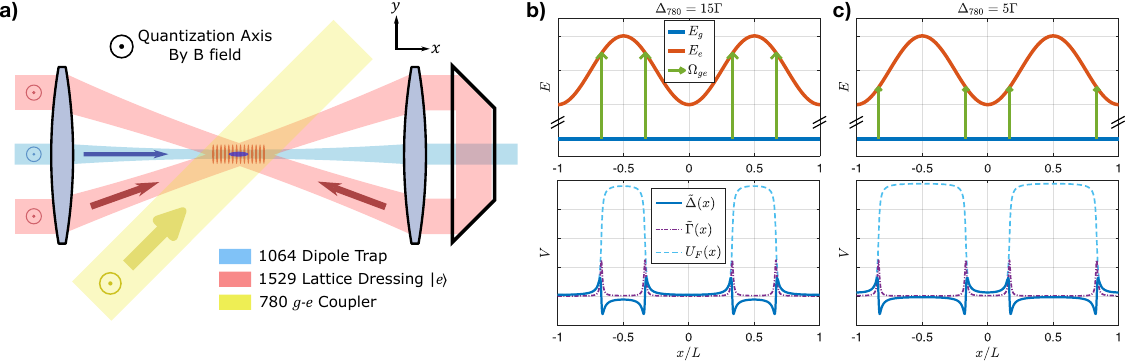}
\caption{\label{fig:U_delta} (a) Optical setup for excited-state engineering. All laser polarizations are aligned with the magnetic field which defines the quantization axis. (b,c) Effective level scheme for periodic excited-state engineering and the corresponding dressing potential $V_\mathrm{eff}$, with $\Delta_\mathrm{780}=15\Gamma$ or $5\Gamma$ for $V_e=20\Gamma$.}
\end{figure*}

The master-equation formalism yields a unified description of the external motion of a driven-dissipative atomic ensemble, capturing both coherent level shifts and dissipation via a complex effective potential $V_\mathrm{eff}(\Delta,\Omega)$. To experimentally benchmark the predicted $V_\mathrm{eff}(\Delta)$ line shape (Fig.~\ref{fig:level}d), a Bose-Einstein condensate (BEC) is exposed to a spatially modulated detuning field while being driven by a uniform laser beam. Details on the creation and caracterization of such spatial detuning can be found in \cite{Maxime_2018,Veyron_2024}. To probe the spatial coherence of the condensate, we modulate the detuning periodically in space, mapping spatial correlations directly into momentum space via Kapitza-Dirac (KD) diffraction. The experimental setup (Fig.~\ref{fig:U_delta}a) comprises the following key components:

\begin{figure}
    \includegraphics[width=0.99\linewidth]{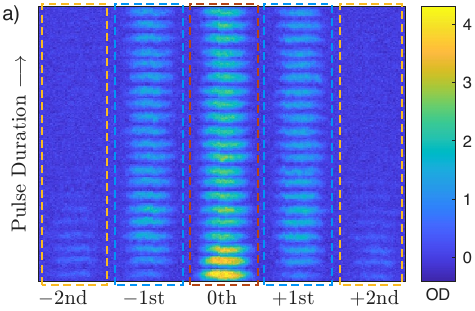}
    \includegraphics[width=0.99\linewidth]{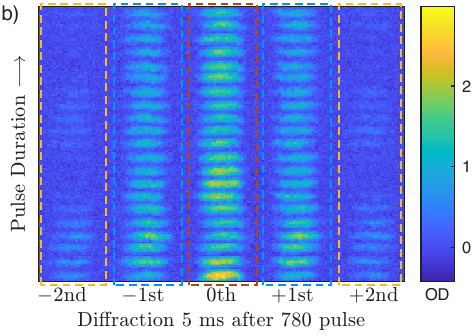}
    \caption{Examples of experimental diffraction patterns versus pulse duration $T$ (scanned increments of $\Delta T=5$ µs, starting from $T=10$ µs at the bottom). (a) $s_\mathrm{780}=0.034$, $\Delta_\mathrm{780}=15\Gamma$, approaching a steady state; (b) $s_\mathrm{780}=0.169$, $\Delta_\mathrm{780}=15\Gamma$, exhibiting persistent oscillations. Red, blue and yellow boxes represent respectively the integration zones of the $0^\mathrm{th}$, $1^\mathrm{st}$ and $2^\mathrm{nd}$ diffraction orders.}
    \label{fig:image_exp}
\end{figure}

\begin{figure*}
    \centering
    \includegraphics[width = 0.99\linewidth]{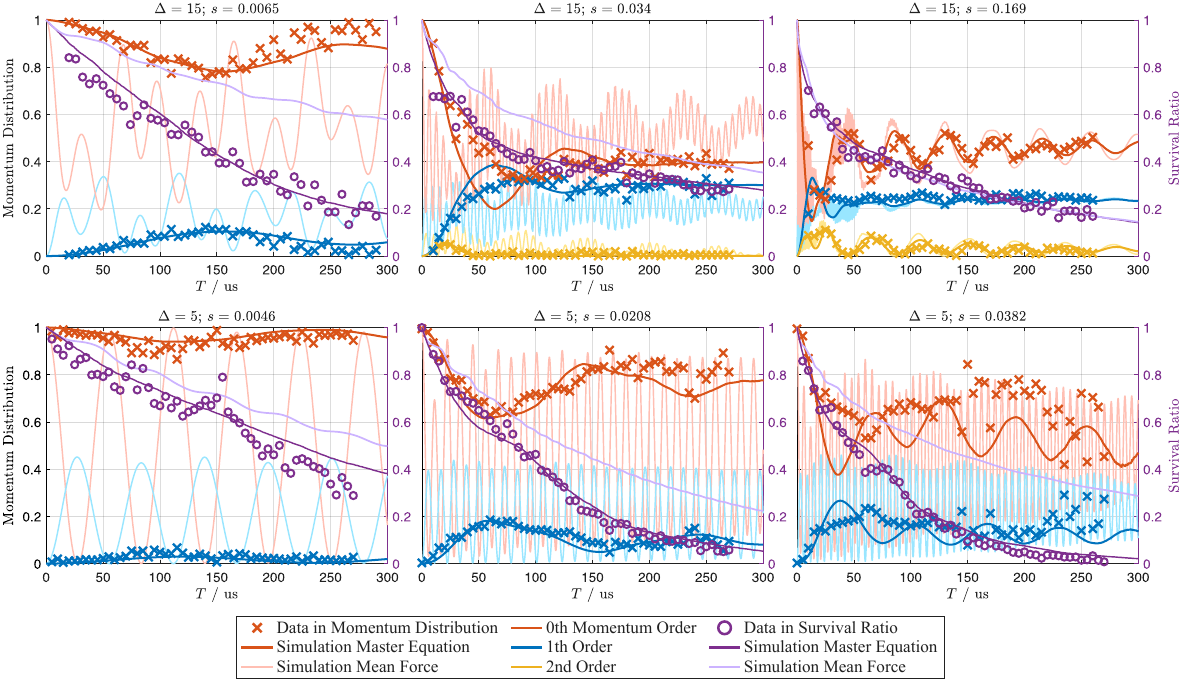}
    \caption{\label{fig:time_evo} Time evolution of the momentum distribution for selected saturation parameters and detunings. Set of parameters are specified above each sub-figure. Experimental diffraction data (markers) are compared with numerical predictions from the mean-force model (light curves) and the effective non-Hermitian master-equation model (solid curves). Red, blue and yellow crosses represent respectively the $0^{th}$, $1^{st}$ and $2^{nd}$ diffraction order population. Purple circle represent the survival ratio which is presented to the right axis. The same color code is used for the corresponding curves of the two theoretical models. For non-zero momentum orders, data are the average of orders with opposite sign.}
\end{figure*}

\begin{itemize}
    \item A pure $^{87}$Rb BEC in the state $\keet{g}=|$5S$_{1/2}$, $F=1,m_F=-1\rangle$,  is produced in a hybrid trap formed by of a 1064 nm  single-beam optical dipole trap (\textcolor{blue}{blue}) focused 100 µm below the center of a quadrupole magnetic trap ($\partial_zB=20$ G/cm). The magnetic field at the atom position sets the quantization axis to vertical. The low atom count ($N\approx5\times10^3$) and tight longitudinal confinement restrict the condensate size to  $w_\mathrm{at}<10$ µm, ensuring that different momentum orders separate clearly within the field of view of our high-NA imaging system. 
    \item A 1529.36 nm optical lattice (\textcolor{red}{red}) is adiabatically loaded after the BEC preparation. This wavelength addresses the $5\mathrm{P}_{3/2}\rightarrow 4\mathrm{D}_{5/2}$ transition of $^{87}\mathrm{Rb}$, generating a periodic level shift for the excited state $|e\rangle = \|5\mathrm{P}_{3/2}, F'=2\rangle$ of the form $U_e(x) = V_e \cos^2(kx)$ with amplitude $V_e = 20\Gamma$ (calibrated in Appendix~\ref{app:calibration}). The residual dipole shift on the ground state is negligible ($V_g \approx 0.5 E_r \sim 10^{-4}\Gamma$). 
    \item A square pulse of a $780~\mathrm{nm}$ laser with beam waist $w_\mathrm{780} = 1~\mathrm{cm} \gg w_\mathrm{at}$ drives near-resonantly with the $g\rightarrow e$ repumping transition. The pulse duration $T$, saturation parameter $s_\mathrm{780}$, and bare-transition detuning $\Delta_\mathrm{780} = \omega_\mathrm{780} - \omega_0$ are the control parameters. For a given $\Delta_\mathrm{780}$, the excited-state modulation $U_e(x)$ induces a spatially periodic detuning $\Delta(x) = \Delta_\mathrm{780} - U_e(x)/\hbar$, yielding a 1D array of effective potentials $V_\mathrm{eff}(x)$ that can be numerically computed using either the local mean force formalism or the master equation approach (Fig.~\ref{fig:U_delta}b,c). The square pulse non-adiabatically projects the initial condensate plane wave onto the Bloch basis of the effective lattice. Sudden pulse turn-off re-projects the atomic state back onto plane-wave states, populating discrete momentum orders $|p_n = \pm 2n\hbar k\rangle$.
    
    \item A $40$ µs blast pulse targeting the $5\mathrm{S}_{1/2}, F=2 \rightarrow 5\mathrm{P}_{3/2}, F'=3$ transition removes the atoms repumped into $F=2$. To further remove $m_F \neq -1$ Zeeman sublevels within $F=1$ manifold, the cloud is held in the hybrid trap for $5~\mathrm{ms}$. During this holding time, momentum orders separate spatially ($2\hbar k t / m \sim 30~\mu\mathrm{m}$) while magnetically untrapped $m_F = 0, +1$ atoms fall away under gravity. Finally, the remaining population in $|g\rangle$ is imaged using a high-numerical-aperture system ($\mathrm{NA}=0.38$, magnification $M=20$; Fig.~\ref{fig:image_exp}).
\end{itemize}

In this report, we focus on two representative detunings $\Delta_\mathrm{780}=15\Gamma$ and $5\Gamma$ (Fig. \ref{fig:U_delta}b\&c). The value of $\Delta_\mathrm{780}$ determines the positions of the resonant coupling, setting both the dissipation peaks position and the boundary between red-detuned ($\tilde{\Delta}(x) < 0$, attractive) and blue-detuned ($\tilde{\Delta}(x) > 0$, repulsive) regions.
Adjusting $\Delta_\mathrm{780}$ relative to the dressing amplitude $V_e$ effectively controls the ratio of length of the two regions. The chosen values of $\Delta_\mathrm{780}$ ($15\Gamma$ and $5\Gamma$) form a pair of reversed ratios for $V_e = 20\Gamma$.

\subsection{Results}


The temporal evolution of the KD diffraction pattern (Fig.~\ref{fig:image_exp}) directly reflects the spatial coherence of the driven-dissipative condensate. By scanning the pulse duration $T$, we reconstruct the time evolution of the distribution in momentum orders for each parameter set $(\Delta_\mathrm{780},s_\mathrm{780})$.

In Fig.~\ref{fig:time_evo}, we present the experimental data and numerical simulation  of the dynamics for six representative parameter configurations ($\Delta_\mathrm{780} = 5\Gamma, 15\Gamma$ across three saturation intensities $s_\mathrm{780}$). The observed behavior follows three distinct regimes where coherent diffraction and dissipation dominate in turn:


\begin{itemize}
    \item At very low saturation, the population exhibits slow, low-amplitude oscillations between the $0^\mathrm{th}$ and $1^\mathrm{st}$ momentum orders, resembling the dynamics of a weakly excited BEC (small population in $1^\mathrm{st}$ excited Bloch state).
    \item At intermediate drive strengths, population transfer to higher momentum states increases, but dissipation plays a crucial role and causes the momentum distribution to damp rapidly toward a steady-state ratio. This is due to the rapid extinction of certain Bloch states, which is the signature of a driven-dissipative dynamics. Another intriguing evidence of the competition between coherent evolution and dissipation is the increase of remaining atoms with increasing drive (from $s=0.0065$ to $s=0.034$ at $\Delta_\mathrm{780}=15\Gamma$).
    \item Above a saturation threshold, at large saturation, robust oscillations re-emerge between momentum orders with higher frequencies and larger amplitudes, indicating that higher excited Bloch states are being populated. The long lifetime of the oscillation indicate that these states are protected from dissipation.
\end{itemize}


Matching the eigenenergy differences of the two model potentials and momentum oscillation frequencies of the data provides a intuitive framework to explain the stark contrast between the master-equation approach and the local mean-force treatment. In Fig.~\ref{fig:time_evo}, the quantitative agreement with experimental data validates the master-equation formalism developed in the preceding section, whereas the local mean-force treatment predicts rapid momentum oscillations that are absent in the experiment.

This discrepancy originates from a fundamental difference in the predicted effective potential depth, illustrated schematically in the lower part of Fig.~\ref{fig:U_delta}b\&c. Integrating the mean force significantly overestimates the potential amplitude, which artificially creates deep potential wells capable of supporting high-energy bound states localized away from resonant dissipation nodes. The absence of these long-lived, high-frequency oscillations in our experimental observations confirms that the mean-force model does not allow to properly render the coherent spatial dynamics of such quantum matter waves in a potential based on detuning modulation.


Having validated the master-equation framework, we turn in the next section to a modal analysis of the non-Hermitian Hamiltonian $H_\mathrm{eff}$. While the direct stochastic Schrödinger simulations in Fig.~\ref{fig:time_evo} yield highly accurate numerical results, they function as a black-box approach that obscures the underlying physical mechanisms. In contrast, decomposing the dynamics into non-Hermitian Bloch modes provides a intuitive physical picture of the interplay between coherent driving and dissipation.

\subsection{Modal analysis}

To gain physical intuition, we will now perform an eigenmode analysis of the effective non-Hermitian periodic Hamiltonian $H_\mathrm{eff}$. Because the full master equation describes a time-irreversible open system — where spontaneous emission causes state purity decay and atom loss — a modal expansion based solely on $H_\mathrm{eff}$ is inherently an approximation that neglects time-irreversible quantum jumps that affect the coherence of the state. Nevertheless, because the internal coherences rapidly relax to their steady state and spatial coherence remains large on the experiment timescales, this non-Hermitian Bloch mode expansion successfully captures both the diffraction dynamics and the atom loss. 
In the following, we will thus compute the complex eigenvalues $E_n$ of $H_\mathrm{eff}$. The real parts $\mathrm{Re}(E_n)$ determine the energy splittings and oscillation frequencies between momentum states, whereas the imaginary parts $\mathrm{Im}(E_n)$ dictate the effective decay rates of the corresponding Bloch modes.


To ease understanding, the modal analysis is organized into two sequential steps. In section~\ref{sec:momentum_diffraction}, we focus on the real parts of the eigen-energies $\mathrm{Re}(E_n)$ and examine the projection of the Bloch modes onto the free-space momentum states, which governs the coherent momentum diffraction patterns. In this initial stage, dissipative losses of each Bloch states are accounted for via their respective decay factor  $e^{\mathrm{Im}[E_n] T / 2 \hbar}$. In section~\ref{sec:drive_dissipation_interplay}, we analyze the parametric trajectories of the complex eigenvalues in the $(\mathrm{Re}[E_n], \mathrm{Im}[E_n])$ plane. This reveals the intrinsic interplay between coherent matterwave diffraction and localized dissipation, providing a physical explanation for the observed atom-number evolution.

\subsubsection{Analyzing coherent momentum diffraction}
\label{sec:momentum_diffraction}

\begin{figure*}
    \centering
    \includegraphics[width=0.95\linewidth]{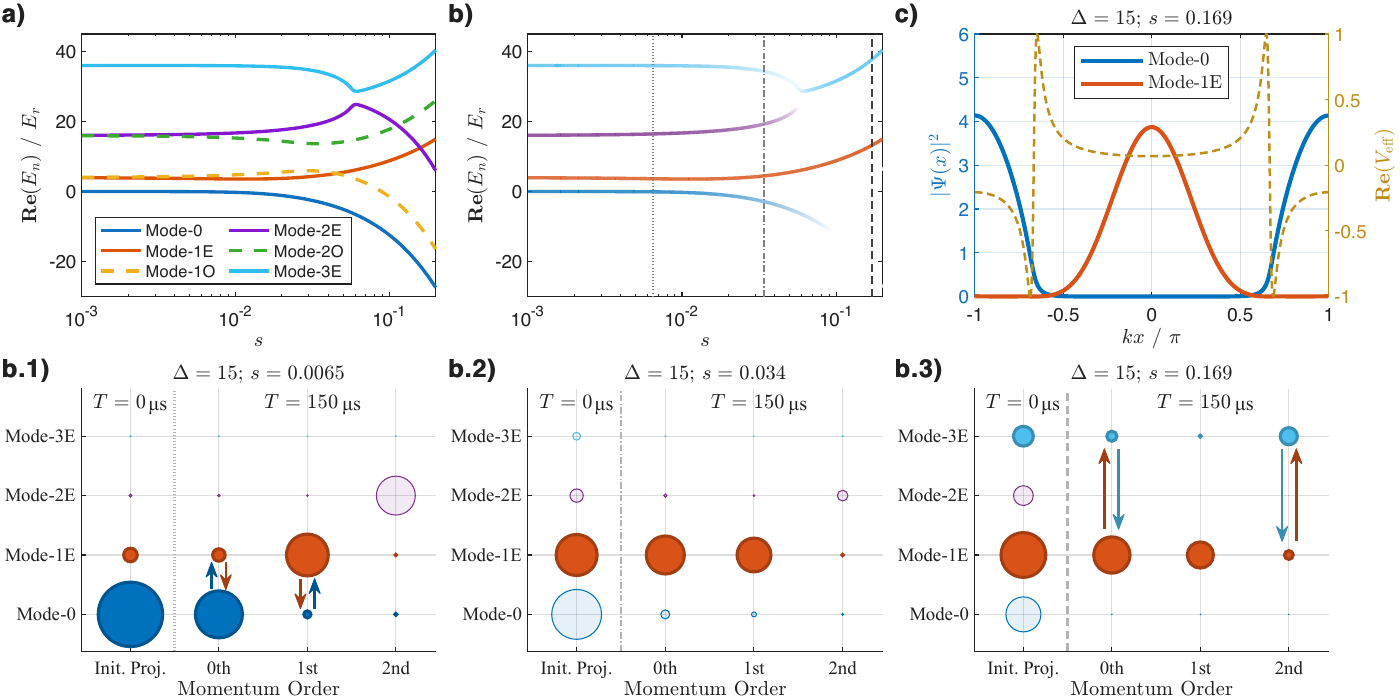}
    \caption{Mode analysis for $\Delta_\mathrm{780}=15\Gamma$: a) $\mathrm{Re(E_n)}$ of eigen modes in the Bloch basis; b) The dominant modes during diffraction indicated by transparency, based on the product of initial projection and remaining fraction at $T=150$ µs; c) The examples of eigen wavefunction for Mode-1E (ground trapping state) and Mode-0 (anti-trapping state) for $s=0.169$; b.1-3) The projection from the Bloch basis onto the free-space momentum, counted at the 3 saturation values indicated on b). The arrows note the momentum order pairs under Rabi oscillation.}
    \label{fig:modes}
\end{figure*}

\begin{figure*}
    \centering
    \includegraphics[width=0.95\linewidth]{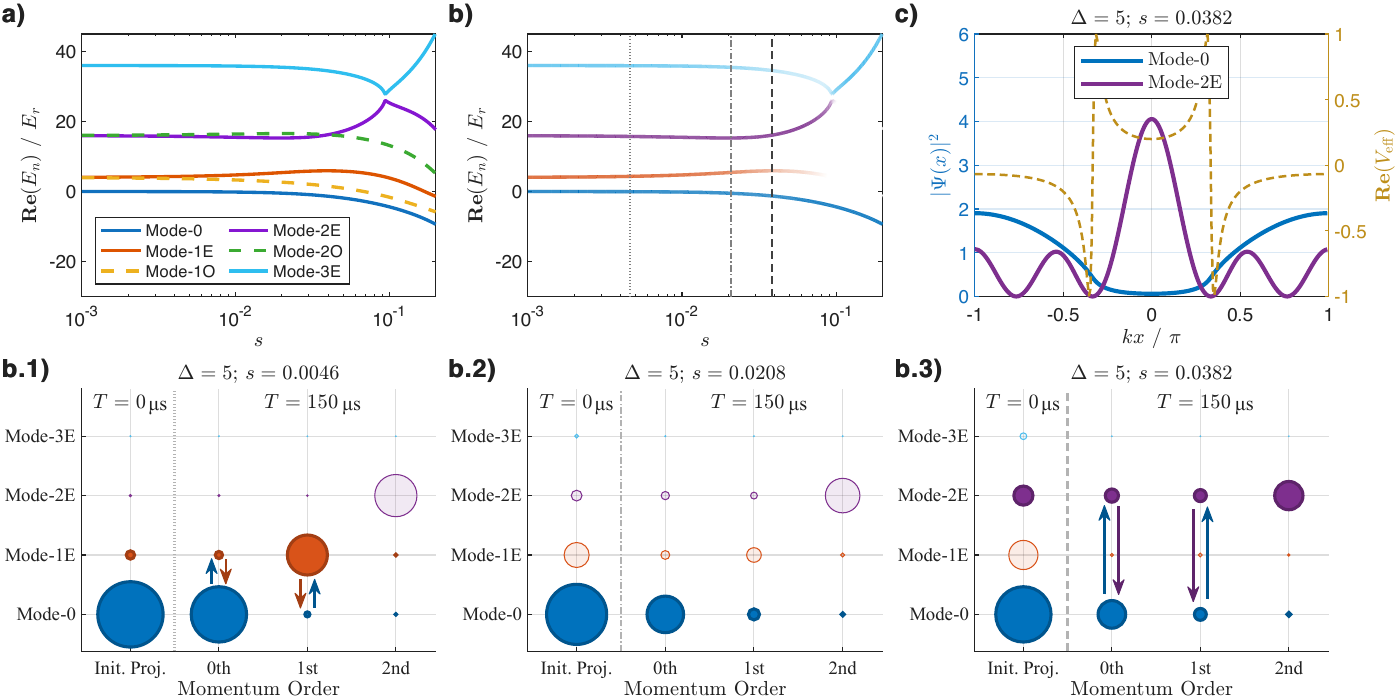}
    \caption{Mode analysis for $\Delta_\mathrm{780}=5\Gamma$ as in Fig. \ref{fig:modes}, except for the values of $s$ specified on the panels b, c and b.1-3.}
    \label{fig:modes5}
\end{figure*}

In Fig. \ref{fig:modes}a, the real parts of the eigenenergies $\mathrm{Re}(E_n)$ are shown as functions of the saturation parameter $s$  for the lowest lying modes that are populated during the experiments. In the weak-pumping limit ($s\rightarrow0$), the potential is flat with periodic boundary conditions. Thus, except for the fundamental plane-wave mode, all excited modes are doubly-degenerate, possessing either symmetric (cosine-like) or anti-symmetric (sine-like) spatial parities.
Based on this asymptotic behavior, we label these modes as Mode-$n$E and Mode-$n$O, where the number $n$ denotes the spatial frequency order, and E/O denotes the parity (Even or Odd). Because our system consists of an initially uniform condensate and a spatially even periodic potential (Fig. \ref{fig:modes}\&\ref{fig:modes5}c), only the coupling to symmetric modes is possible. As a result, the anti-symmetric Mode-$n$O (plotted in dashed lines) have zero overlap with the $0^\mathrm{th}$ order plane wave during the initial projection onto the Bloch basis, and are thus not involved in the diffraction dynamics. 

At large saturation ($s\approx10^{-1}$), a different asymptotic behavior emerges: $\mathrm{Re}(E_n)$ scales  linearly with the saturation parameter $s$, dividing the eigenmodes into two distinct families. Taking the six lowest modes at $\Delta = 15\Gamma$ as an example (Fig. \ref{fig:modes}a), half of them exhibit negative linear slopes while the other half display positive slopes of equal magnitude. Examining the corresponding wavefunction profiles in Fig.~\ref{fig:modes}c reveals that these two families map to modes localized either in the blue-detuned region (trapping modes) or in the red-detuned region (anti-trapping modes). When the wavefunction becomes tightly confined within one of these two regions at large $s$, this linear shift reflects the local potential offset of the blue- or red-detuned regions relative to $E=0$. Note that the $s \rightarrow 0$ modes do not map one-to-one onto the large-$s$ asymptotic states — as evidenced by the anti-crossing between Mode-2E and Mode-3E — because the Bloch wavefunction width reduces as $s$ increases, while the widths of free sinusoidal functions at $s=0$ are constant. To avoid ambiguity, in the following, we adopt the  Mode-$n$E/O labels as defined at $s=0$. We also emphasize that the two asymptotic basis are adiabatically connected by a change of $s$.


In addition to the eigenmode structure, the initial condition plays a crucial role in the upgoing dynamics. The initial population of each Bloch modes is set by the overlap with the plane wave BEC. Unlike unitary Hermitian systems, different eigenmodes in a driven-dissipative system may experience vastly different decay rates.
To capture this effect, Fig.~\ref{fig:modes}b displays $\mathrm{Re}[E_n(s)]$ with line transparencies weighted by the product of the initial population overlap and the survival factor $\exp[\mathrm{Im}(E_n)T/2]$. This representation reveals three distinct saturation regimes, each characterized by a different dominant, long-lived mode structure. Combined with the momentum decomposition of these modes (Figs.~\ref{fig:modes}b. 1-3 and ~\ref{fig:modes5}b. 1-3), the observed diffraction patterns can be qualitatively interpreted. As an example, we now develop the mode analysis for $\Delta_\mathrm{780}=15\Gamma$ (Fig. \ref{fig:modes}):

\begin{itemize}
    \item At low saturation, the first excited Mode-1E is gradually populated as the saturation increases, leading to a Rabi oscillation of a small amplitude between the 0th and 1st diffraction orders. As Mode-0E has relatively larger loss rate, dissipation tends to equalize the populations of the two modes, resulting in an increasing constrast of the oscillations with time.
    \item The intermediate regime is the simplest case. Although the population of the ground Mode-0E is reduced, it is still comparable to the Mode-1E at $T=0$, leading to a pronounced initial oscillations. However, because Mode-0E is positioned in the red-detuned region and decays rapidly, these oscillations damp quickly, leaving a steady-state population in Mode-1E whose momentum distribution matches the projection in Fig.~\ref{fig:modes}b2.
    \item For the high saturation regime, the Mode-0E still has a non-negligible initial population, but its decay rate is very large. The dominant long-living modes are then Mode-1E and Mode-3E, both localized in the blue-detuned region. Interference between these two modes drives robust oscillations between the $0^\mathrm{th}$ and $2^\mathrm{nd}$ momentum orders while leaving the $1^\mathrm{st}$ order nearly constant, as Mode-3E has negligible overlap with the $1\mathrm{st}$ momentum order. Such a property is determined geometrically by the ratio of length between the blue-detuned region and the lattice period ($\Delta_\mathrm{780}=15\Gamma$). The larger energy difference between Mode-1E and Mode-3E yields a significantly higher oscillation frequency than in the low-saturation regime.
\end{itemize}

This modal analysis provides an effective understanding of the observed diffraction dynamics. The same investigation of the dominant modes contribution explain the behavior observed for $\Delta_\mathrm{780}=5\Gamma$ (Fig. \ref{fig:modes5}b). Here, the asymptotic behaviors at large $s$ reveals that the lowest-lying modes are all anti-trapping states subjected to strong losses. Because higher-order anti-trapping modes decay even faster, the ground Mode-0E remains the dominant mode throughout the saturation range we investigate. The oscillations in the diffraction pattern are then determined by which secondary mode pairs with Mode-0E. At $s\rightarrow0$, where the lattice geometry imposes the least effect, weak lattice modulation leads to low-amplitude Rabi oscillations with Mode-1E. 
At larger saturation ($s=0.0382$), the pairing mode shifts to Mode-2E, which is centered in the blue-detuned region (Fig. \ref{fig:modes5}c) and thus long-living, giving rise to the rapid oscillations observed experimentally. Between these limits ($s=0.0208$),  three modes are involved with the Mode-0E being the dominant. Because their energy splittings are not commensurate, destructive interference reduces the observed net oscillation amplitude.


Comparing Figs.~\ref{fig:modes}c and \ref{fig:modes5}c highlights geometry, and its dependence on $\Delta_\mathrm{780}$ as the primary governing factor.
For $\Delta_\mathrm{780}=5\Gamma$, the blue-detuned trapping area is so narrow that even the fundamental trapped mode has a spatial width comparable to a high frequency plane wave. This mode structure is directly analogous to electromagnetic wave propagation in an optical waveguide, where the saturation parameter $s$ controls the core-cladding refractive index contrast and the blue- and red-detuned regions act as core and cladding, respectively. For a narrow core ($\Delta_\mathrm{780} = 5\Gamma$), no guided (long-lived) modes are supported below a critical index contrast (saturation). While for a larger core ($\Delta_\mathrm{780}=15\Gamma$), the waveguide evolves from a mono-mode ($s=0.034$) to a multi-mode one ($s=0.169$) when the core-cladding refractive index contrast increases (increasing $s$). Still, for mode engineering in the waveguide, we generally consider the materials as uniform (constraint by real-life material engineering), while this detuning modulated 'matter waveguide' has an extraordinary profile $V_\mathrm{eff}(x)$. This non-uniformity along with the dissipation concentrated at the boundary introduces extra drive-dissipation interplay presented in the next section.

\begin{figure*}
    \centering
    \includegraphics[width=0.96\linewidth]{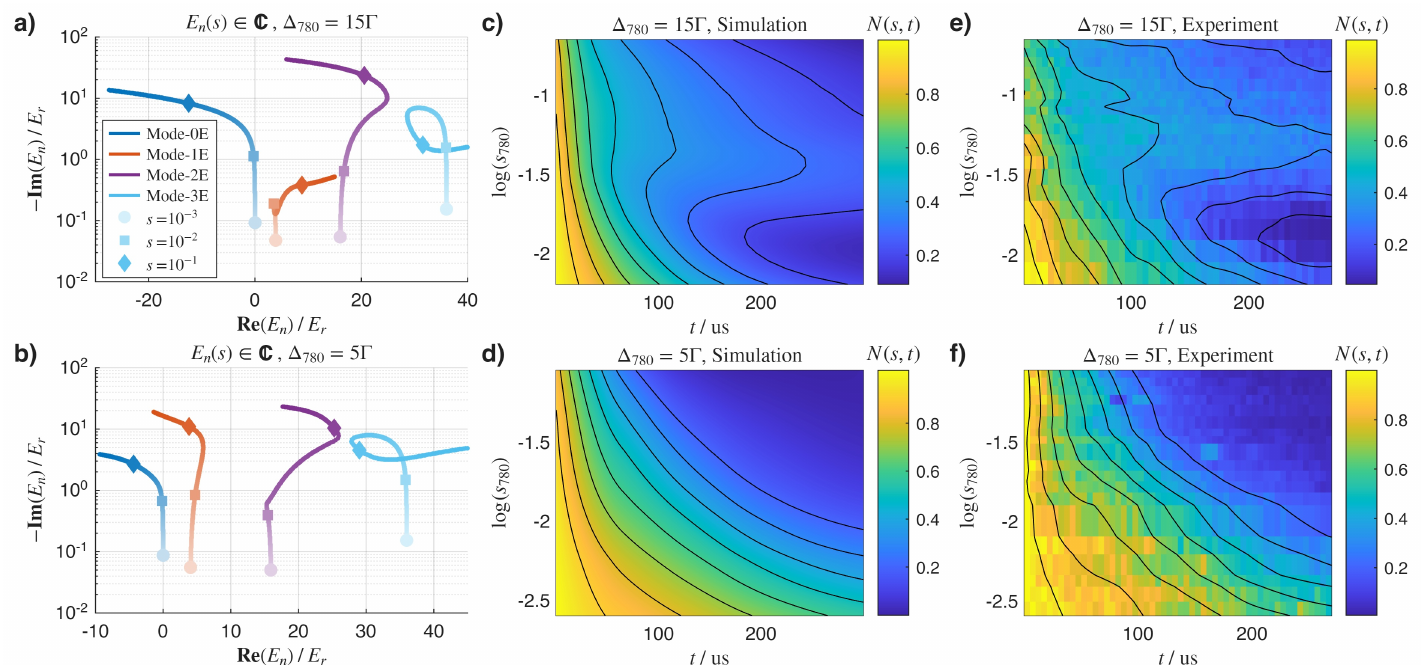}
    \caption{a,b) The parametric trajectories of the complex eigen energies $E_n$ as the functions of the saturation parameter $s$, for the even modes involved in the experiments. The imaginary part is taken reverse sign to match positively the loss. The increasing opacity represents increasing saturation and the markers represent the saturation thresholds $s=10^{-3},~10^{-2},~10^{-1}$. c,d) Numerical simulation of remaining atom fraction $N(s,t)$ based on the stochastic \sch{} equation; e,f) Experimental measurement of $N(s,t)$.}
    \label{fig:Nat_2dmap}
\end{figure*}

\subsubsection{Interplay between the coherent drive and dissipation}
\label{sec:drive_dissipation_interplay}

In our experiment, the spatial modulation of the detuning $\Delta(x)$ is realized through excited state engineering \cite{Maxime_2018,Veyron_2024}. The detuning dependence of the potential $V_\mathrm{eff}(\Delta)$ derived in section \ref{sec:th}, allows to obtain the spatially varying effective potential. Constrained by the Kramers-Kronig relations, the minimum of the coherent light shift (trap center) cannot coincide with the maximum of the dissipation rate. In particular, when atoms are confined in blue-detuned regions, the potential minimum coincides with a local minimum in dissipation. This spatial offset produces an interesting interplay between coherent trapping and dissipation, visualized via the parametric trajectories of the eigen energies [$\mathrm{Re}(E_n),\mathrm{Im}(E_n)$] in the complex plane in Fig. \ref{fig:Nat_2dmap}a\&b. Choice of coordinates is such that loss increase upwards ($-\mathrm{Im}(E_n)\rightarrow + \infty$). The trajectories are parametrized by $s$ which is encoded in the transparency of the curve ($s=0 \Leftrightarrow $ "Fully transparent"). We will now interpret these trajectories, which present different asymptotic behaviors, for the trapping or anti-trapping modes:

\begin{itemize}
    \item In the weak-pumping limit $s\rightarrow0$, the eigen wavefunction is shaped as sinusoidal functions in a flat periodic potential (of period $L_{\rm lat}$), and does not favor a specific region. The coherent level shift averages approximately to zero over each lattice period. In other word, the real part of the Eigenenergies ($\mathrm{Re}(E_n)$) are given by their kinetic energy term $n \times \frac{h^2}{2 m L_{\rm lat}^2}$ while their loss rate $\mathrm{Im}(E_n)$, that originates from scattering, scales linearly with $s$. The resulting parametric traces are thus almost vertical in this regime ($s<10^{-3}$).
    \item As the saturation increases, the Eigen wavefunctions become progressively confined within either the blue- or red-detuned regions, and subsequently away from the resonant dissipation nodes. For blue-detuned trapping modes (ex: \textcolor{red}{red} Mode-1E as in Fig. \ref{fig:modes}c), this spatial contraction causes an anomalous \textit{decrease} in decay rate for \textit{increasing} drive intensity. Because the ensemble-averaged loss rate should scale linearly with saturation, these long-lived modes are effectively "protected" by the rapidly decaying anti-trapping states. For instance, in Figs.~\ref{fig:Nat_2dmap}a and \ref{fig:Nat_2dmap}b, the modes highlighted in \textcolor{purple}{purple} (Mode-2E) exhibit an increase in loss rate larger than one order of magnitude while $s$ increase by exactly one order from $10^{-2}$ to $10^{-1}$.
    \item Beyond this intermediate regime, the loss rate recovers the ordinary behavior and positively increase with the saturation for $s\gtrsim10^{-1}$. Physically, when the eigen wavefunctions are already localized in either region, further reduction in width cannot increase significantly the lifetime. At this stage, the off-resonant scattering becomes the dominant loss mechanism, causing the mode decay rates to scale positively with $s$ once again.
\end{itemize}

This protection effect of the intermediate regime manifests experimentally in the remaining atom fraction $N(s,t)$ as a function of the saturation parameter $s$ and the pulse duration $t$. Figures \ref{fig:Nat_2dmap}c-f show 2D maps of $N(s,t)$ comparing measurements (panel e-f) with numerical simulations (panel c-d).
For $\Delta_\mathrm{780} = 15\Gamma$, this anomalous protection produces a distinct "dip" in atom loss at intermediate saturation for long holding times $T$: as $T$ increases, the dominant mode is the most long-living Mode-1E which is centered in the blue-detuned region and whose decay rate presents a local minimum at $s\approx10^{-2}$.
For $\Delta_\mathrm{780} = 5\Gamma$, although trapping modes are never the dominant ones, the spatial mismatch between peak dissipation and the potential minimum still induces a clear change of slope in the equi-contours of Fig. \ref{fig:Nat_2dmap}c-f. The competition between coherent trapping and dissipation is a universal feature of detuning-engineered light-matter interactions that could be further investigated on other detuning profiles that highlight their interplay in various configurations.

\section{Conclusion and perspectives}\label{sec:persp}

We have presented a master-equation framework to describe the non-Hermitian dynamics of a two-level quantum-degenerate matter wave driven by a near-resonant laser field. This non-perturbative approach provides a unified picture of both coherent light shifts and dissipation, bridging the gap between the far-detuned and near-resonant regimes. The formalism extends naturally to open multi-level systems or to the strong drive limit, where higher order corrections are cross-validated by the resolvent formalism (Appendix \ref{app:resolvent}). Experimentally, we benchmarked these non-Hermitian dynamics by diffracting a condensate off an engineered optical grating. The quantitative agreement between experimental data and master-equation simulations confirms the validity of our approach over conventional mean-force models. Furthermore, complex eigenmode analysis provided an intuitive understanding for the consequences of the $V_\mathrm{eff}(\Delta)$ profile and the resulted interplay between the coherent drive and dissipation, while the experimetal realization serves as a practical showcase of how the frequency response can be imprinted onto the spatial-modulation. 

This work establishes a rigorous foundation for controlling driven-dissipative quantum systems via light-matter interactions. By quantifying interactions in the near-resonant regime, it expands the quantum control toolkit to harness strong coherent dynamics alongside dissipation. The steep potential gradients near atomic resonances offer a pathway toward generating ultra-high-frequency dressing potentials ($\gg \lambda^{-1}$), accelerating state manipulation to timescales approaching $\Gamma^{-1}$. Additionally, the spatial separation of real and imaginary response peaks allows one to pre-shape atomic wavefunctions into protected, long-lived modes for robust state engineering. Beyond single-particle control, this well-calibrated complex potential provides a ideal testing ground for strongly correlated driven-dissipative dynamics, where interatomic interactions would play a central role.

\appendix
\begin{widetext}
\section*{Appendix}\nonumber
\subsection{Calculation of Raman decay}\label{app:raman}

The difference with the two-level system comes from the loss of the total atom number, which leads to the re-definition of local density matrix $\rho$:
\begin{equation}
    \rho(t)=\rho^R(t)e^{-\kappa t}
\end{equation}

We will now determine the local steady states of the fast-damping terms $\rho_{ge,eg,ee}^{R\mathrm{,st}}$ and start with the coherence terms $\rho_{ge,eg}^{R\mathrm{,st}}$:

\begin{equation}\label{eq:SteadyImpureRaman}
    \begin{aligned}
    &\partial_t\rho_{ge}^R - \kappa\rho_{ge}^R = 
    - \rho_{ge}^R\left(\frac{\Gamma}{2}+i\hat\Delta\right)
    - \frac{i}{2}\left(
    \hat\Omega^*\rho_{ee}^R
    -\rho_{gg}^R\hat\Omega^*\right) \\
    &\partial_t\rho_{eg}^R - 
    \kappa \rho_{eg}^R = 
    - \left(\frac{\Gamma}{2}-i\hat\Delta\right)\rho_{eg}^R
    - \frac{i}{2}\left(\hat\Omega\rho_{gg}^R-
    \rho_{ee}^R\hat\Omega\right)
    \end{aligned}
\end{equation}
in which we expand $\partial_t(\rho^Re^{-\kappa t}) = e^{-\kappa t}(\partial_t-\kappa)\rho^R$ and remove the common $e^{-\kappa t}$ terms. Substituting the condition $\partial_t\rho^R=0$, we obtain the steady values $\rho_{ge,eg}^{R\mathrm{,st}}$:

\begin{equation}
    \begin{aligned}
    &\rho_{ge}^{R,\mathrm{st}}=
    - \frac{i}{2}\left(\hat\Omega^*\rho_{ee}-\rho_{gg}\hat\Omega^*\right)
    \left[\left(\frac{\Gamma}{2}-\kappa\right)+i\hat\Delta\right]^{-1}\\
    &\rho_{eg}^{R,\mathrm{st}}= 
    - \frac{i}{2}\left[\left(\frac{\Gamma}{2}-\kappa\right)-i\hat\Delta\right]^{-1}
    \left(\hat\Omega\rho_{gg}-\rho_{ee}\hat\Omega\right)
    \end{aligned}
\end{equation}

Substituting the local steady solution back into the equation of $\hat\rho_{gg}$, we obtain:

\begin{equation}\label{eq:ddp_ptrg_R}
\begin{split}
    \partial_t\hat\rho_{gg}^R =
    - \frac{i}{\hbar}\left\{
    \frac{\hbar|\Omega|^2\left[\hat\Delta-i(\Gamma-2\kappa)/2\right]}
    {(\Gamma-2\kappa)^2+4\hat\Delta^2}
    (\hat\rho_{gg}^R-\hat\rho_{ee}^R) -
    (\hat\rho_{gg}^R-\hat\rho_{ee}^R) 
    \frac{\hbar|\Omega|^2
    \left[\hat\Delta+i(\Gamma-2\kappa)/2\right]}
    {(\Gamma-2\kappa)^2+4\hat\Delta^2}
    \right\}\\
    -\frac{i}{\hbar}\left[\frac{p^2}{2m},\hat\rho_{gg}^R\right] +\kappa\hat\rho_{gg}^R+(1-a)\Gamma\hat\rho_{ee}^R
\end{split}
\end{equation}

Since we have separated the exponential loss $e^{-\kappa t}$, the remaining term $\hat\rho^R$ is trace-conserving. Thus all the real terms on the right side should sum up to 0, leading to the following constraint for the local excited population $\rho_{ee}^R$:

\begin{equation}
    \mathbf{Re}(\partial_t\rho_{gg}^R)=-\frac{|\Omega|^2(\Gamma-2\kappa)}
    {(\Gamma-2\kappa)^2+4\hat\Delta^2}
    (\rho_{gg}^R-\rho_{ee}^R)+\kappa\rho_{gg}^R+(1-a)\Gamma\rho_{ee}^R = 0.
\end{equation}
We define the shifted saturation parameter $s^R$ as
\begin{equation}
    s^R = \frac{2|\Omega|^2}
    {(\Gamma-2\kappa)^2+4\Delta^2}
\end{equation}
and replace $\rho_{gg}^R = 1-\rho_{ee}^R$, the equation solves the local steady value $\rho_{ee}^{R,\mathrm{st}}$:
\begin{equation}
    \rho_{ee}^{R,\mathrm{st}} = \frac{1}{2}
    \frac{s^R}{\left(1+\frac{\kappa}{\Gamma-2\kappa}\right)+s^R}
\end{equation}
which reduces to the two-level value $\rho_{ee}^\mathrm{st}=s/[2(1+s)]$ for $a=\kappa=0$. Together with $\kappa = a\Gamma\rho_{ee}^{R,\mathrm{st}}$, this self-consistent system determines the higher-order corrections for open systems.

\subsection{Cross-validation from the resolvent formalism for pure Raman decay under the local approximation}\label{app:resolvent}

Across the entire saturation range examined in this work, this condition is well satisfied: the characteristic timescale of momentum excitation exceeds $1~\mu\mathrm{s}$, whereas the spontaneous decay lifetime is $\Gamma^{-1} \approx 25~\mathrm{ns}$.

We revisit the master equation approach to identify the physical regime over which the formalism is applicable. Our derivation starts from Eq. \ref{eq:obeSpace}, in which the spatial distribution of radiation vacuum modes was neglected. In free space, this distribution is homogeneous and the corresponding net recoil momentum averages to zero. Thus, the zero-recoil approximation shares the same foundational condition as the adiabatic approximation: all fluctuations induced by spontaneous decay occur on the timescale of $\Gamma^{-1}$, which must be compared with the characteristic timescale of external motion.
In our experiment, this condition can be expressed via the evolution rate of the state expanded in the discrete momentum basis $|\Psi\rangle = \sum_n c_n |\pm 2n \hbar k\rangle$:

\begin{equation}
    \frac{dc_n}{dt}\ll\Gamma
\end{equation}
Across the entire saturation range examined in this work, this condition is well satisfied: the characteristic timescale of momentum excitation exceeds $1$ µs, whereas the spontaneous decay lifetime is $\Gamma^{-1} \approx 25~\mathrm{ns}$.

The second approximation adopted in our model is the localization approximation applied to higher-order corrections beyond the first-order potential $V_\mathrm{eff}^{(1)}$, yielding the \SB{two}-level $V_\mathrm{eff}$ and the multi-level Raman corrected potential $V_\mathrm{eff}^R$. This derivation can be independently cross-validated in the specific limit of pure Raman decay, where all decay channels lead out of the primary two-level system ($a = 1$; see Fig.~\ref{fig:level}b). In this limit, the ground state level shift $E_g$ can be derived exactly using the resolvent formalism \cite{steck_qao}:

\begin{equation}
    E_g=\frac{\hbar}{2}\left[
    \left(-\Delta-\frac{i\Gamma}{2}\right)-
    \sqrt{
    \left(-\Delta-\frac{i\Gamma}{2}\right)^2+\Omega^2
    }~
    \right]
\end{equation}
Because the resolvent formalism provides a local solution based solely on the internal dynamics, it can be directly compared to $V_\mathrm{eff}^R$ in Eq. \ref{eq:veff_r}. For this purpose, we first expand $E_g$ up to the 2nd order of saturation parameter $s/2=\Omega^2/(\Gamma^2+4\Delta^2)$, yielding:

\begin{equation}
\begin{aligned}
    E_g&=\frac{\hbar}{2}\left[
    \left(-\Delta-\frac{i\Gamma}{2}\right)-
    \left(-\Delta-\frac{i\Gamma}{2}\right)
    \sqrt{
    1+\frac{\Omega^2}{\left(\Delta+i\Gamma/2\right)^2}
    }~
    \right] \\
    &\approx \frac{\hbar}{2}\left[
    \left(-\Delta-\frac{i\Gamma}{2}\right)
    \left(\frac{1}{2}\frac{\Omega^2}{\left(\Delta+i\Gamma/2\right)^2}
    -\frac{1}{8}\frac{\Omega^4}{\left(\Delta+i\Gamma/2\right)^4}
    \right)~
    \right]\\
    &=\frac{\hbar\Omega^2\left(\Delta-i\Gamma/2\right)}{\Gamma^2+4\Delta^2}\left[
    1
    -\frac{1}{4}\frac{\Omega^2}{\left(\Delta+i\Gamma/2\right)^2}
    \right]\\
    &=    \frac{\hbar\Omega^2\left(\Delta-i\Gamma/2\right)}{\Gamma^2+4\Delta^2}\left[
    1
    -\frac{1}{4}\frac{\Omega^2(\Delta^2-\Gamma^2/4-i\Gamma\Delta)}{\left(\Delta^2+\Gamma^2/4\right)^2}
    \right]\\
    &=\frac{\hbar\Omega^2\left(\Delta-i\Gamma/2\right)}{\Gamma^2+4\Delta^2}\left[
    1-\frac{\Omega^2}{\Gamma^2+4\Delta^2}+\frac{\Omega^2}{\Gamma^2+4\Delta^2}\cdot \frac{\Gamma^2/2+i\Gamma\Delta}{\Gamma^2/4+\Delta^2}
    \right]
\end{aligned}
\end{equation}
The first order is exactly $V_\mathrm{eff}^{(1)}$ in Eq. \ref{eq:veff_1st} derived from the master equation. For the second order pure Raman decay, we write respectively the real and imaginary parts:

\begin{equation}
\begin{aligned}
    \mathbf{Re}(E_g) &= \frac{\hbar\Omega^2}{\Gamma^2+4\Delta^2}\left[
    1-\frac{\Omega^2}{\Gamma^2+4\Delta^2}+\frac{\Omega^2}{\Gamma^2+4\Delta^2}\cdot
    \frac{4\Gamma^2}{\Gamma^2+4\Delta^2}
    \right]\cdot \Delta\\
    \mathbf{Im}(E_g) &= \frac{\hbar\Omega^2}{\Gamma^2+4\Delta^2}\left[
    1-\frac{3\Omega^2}{\Gamma^2+4\Delta^2}+\frac{\Omega^2}{\Gamma^2+4\Delta^2}\cdot
    \frac{4\Gamma^2}{\Gamma^2+4\Delta^2}
    \right]\cdot \frac{-i\Gamma}{2}
\end{aligned}
\end{equation}

Then, we expand $V_\mathrm{eff}^{R}$ in Eq. \ref{eq:veff_r} for branching ratio $a=1$ (pure Raman decay). Under such a condition, $\kappa/\Gamma = \rho_{ee}$ and $\rho_{ee}^{(1)}=\Omega^2/(\Gamma^2+4\Delta^2)$. The second order correction for $V_\mathrm{eff}^{R}$ is then calculated at first order in $\kappa$ and $\rho_{ee}$:

\begin{equation}
\begin{aligned}
    V_\mathrm{eff}^R&=
    \frac{\hbar\Omega^2[\Delta-i(\Gamma-2\kappa)/2]}{(\Gamma-2\Gamma\rho_{ee})^2+4\Delta^2}
    \cdot (1-\frac{\rho_{ee}^\mathrm{st}}{\rho_{gg}^\mathrm{st}})\\
    &\approx\frac{\hbar\Omega^2[\Delta-i\Gamma/2\cdot(1-2\rho_{ee})]}{\Gamma^2+4\Delta^2-4\Gamma^2\rho_{ee}}\cdot(1-\rho_{ee})\\
    &\approx
    \frac{\hbar\Omega^2[\Delta-i\Gamma/2\cdot(1-2\rho_{ee})]}{\Gamma^2+4\Delta^2}
    \cdot 
    \left(1+\frac{4\Gamma^2\rho_{ee}}{\Gamma^2+4\Delta^2}\right)
    \cdot(1-\rho_{ee})\\
    &\approx
    \frac{\hbar\Omega^2[\Delta-i\Gamma/2\cdot(1-2\rho_{ee})]}{\Gamma^2+4\Delta^2}
    \cdot 
    \left(1
    -\frac{\Omega^2}{\Gamma^2+4\Delta^2}
    +\frac{\Omega^2}{\Gamma^2+4\Delta^2}\cdot \frac{4\Gamma^2}{\Gamma^2+4\Delta^2}\right)
\end{aligned}
\end{equation}

Expanding $(1 - 2\rho_{ee}^{(1)})$ to first order in saturation shows that the real and imaginary parts of $V_\mathrm{eff}^R$ match the resolvent expressions. This rigorous equivalence up to second order demonstrates that the master-equation approach combined with the localization approximation coincides with the \textit{ab initio} local resolvent formalism. This agreement confirms the mathematical consistency of our localization procedure in Appendix~\ref{app:raman} and corroborates the validity of $V_\mathrm{eff}^R$ across the full range of branching loss ratios $a \in [0, 1]$.

\end{widetext}

\subsection{Numerical simulation method}\label{app:numerical}

\begin{figure*}
    \centering
    \includegraphics[width=0.99\linewidth]{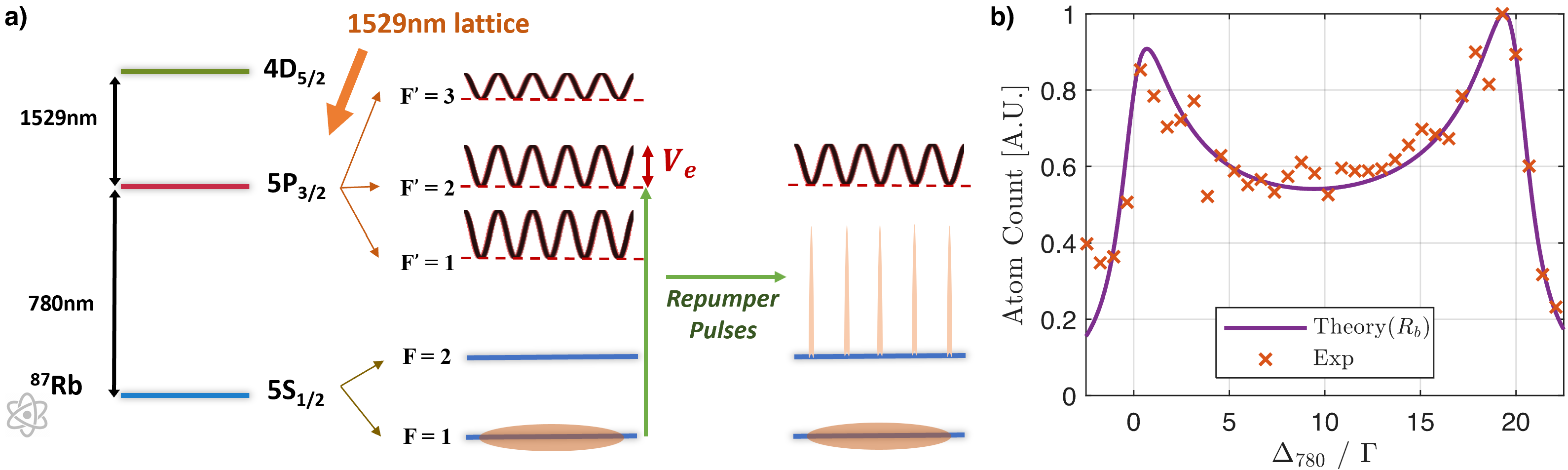}
    \caption{a) Experimental scheme of excited state modulation calibration; b) Measured atom number $N(\Delta_\mathrm{780})$ after short repumper pulses ($s\approx0.05,~T=2$ µs), as a function of repumper detuning $\Delta_\mathrm{780}$ with respect to the bare transition. The fit is based on the Eq. \ref{eq:N_R}\&\ref{eq:localSponRate_broad}, which yield the modulation amplitude $V_e\approx20.1\Gamma$.}
    \label{fig:bat_scheme}
\end{figure*}

To numerically solve the master equation, we adopt the quantum jump algorithm. The principle is to consider the density matrix as statistics: $\rho=\sum P_{\psi}\keet{\psi}\brra{\psi}$, and use Monte-Carlo method to generate a large number of samples that comply with the probability distribution. 

We follow the derivation in \cite{MCWF_dalibard}. For any pure state $\keet{\psi}$, any time update leads to 2 possibilities:

\begin{itemize}
    \item the state evolves coherently under the effective Hamiltonian $H_\mathrm{eff}$: $\keet{\psi(t+dt)} = \mu(1-iH_\mathrm{eff}dt/\hbar)\keet{\psi(t)}$, where $\mu$ represents the normalization coefficient, as the non-Hermitian Hamiltonian on its own does not conserve the trace.
    \item the state undergoes the spontaneous decay and resets to the ground state regardless of its original state: $\keet{\psi(t+dt)}=\keet{g}$.
\end{itemize}

The probability of a quantum jump occurring during $dt$ is $dp = \Gamma \rho_{ee} dt$. Reconstructing the updated density matrix as the weighted sum of both outcomes yields:

\begin{widetext}
\begin{equation}
\begin{aligned}
    \rho(t+dt)&=\mu^2(1-dp)(1-iH_\mathrm{eff}dt/\hbar)\keet{\psi(t)}\brra{\psi(t)}(1+iH_\mathrm{eff}^\dagger dt/\hbar)+dp\keet{g}\brra
    g \\
    &\simeq \mu^2(1-dp)\left(\rho(t)-i[H_\mathrm{eff},\rho]dt/\hbar\right)+dp\keet{g}\brra
    g
\end{aligned}
\end{equation}
\end{widetext}

Expressing the projection operator as $|g\rangle\langle g| = \sigma \rho \sigma^\dagger / \rho_{ee}$ and substituting $\mu^2 (1 - dp) = 1$, the recurrence relation recovers the Lindblad master equation:


\begin{equation}
    \frac{\rho(t+dt)-\rho(t)}{dt}=-\frac{i}{\hbar}[H_\mathrm{eff},\rho]+\Gamma\sigma\rho\sigma^\dagger
\end{equation}

Note that the normalization relation writes $\mu^2=(1-dp)^{-1}$, which indicates that without normalization, the reduction of trace due to the non-Hermitian Hamiltonian is exactly the integral over the jump probability:

\begin{equation}
    1-|H_\mathrm{eff}\keet{\psi(t)}|^2 = \int dp
\end{equation}

This relation defines the cumulative distribution function for the time delay $t$ between successive quantum jumps. Rather than propagating state vectors across fixed, small time increments $\delta t$, we sample a sequence of jump timestamps $\{t_n\}$ directly from this distribution. Between jumps ($t \in [t_n, t_{n+1})$), the state vector evolves deterministically along a normalized trajectory:

\begin{equation}
    \keet{\psi(t)}=\frac{e^{-iH_\mathrm{eff}(t-t_n)}\keet{\psi_0=g}}{||e^{-iH_\mathrm{eff}(t-t_n)}\keet{\psi_0=g}||}
\end{equation}
For non-closed transitions involving Raman decay into reservoir states $\{|R\rangle\}$ (Fig.~\ref{fig:level}b), a quantum jump resets the state vector to $(1-a)^{1/2}|g\rangle$ (rather than $\keet{g})$, where the reduced norm directly reflects irreversible atom loss.

In this work, spatial continuous dynamics are simulated under an effective 1D spatial Hamiltonian:

\begin{equation}
    H_\mathrm{eff} = \frac{p^2}{2m}+V_\mathrm{eff}(x)
\end{equation}
where $V_\mathrm{eff}(x)$ is either:

\begin{itemize}
    \item Eq. \ref{eq:veff_r} derived from the master equation approach;
    \item Or, the mean force integration Eq. \ref{eq:U_force} for the coherent level shift plus the loss: 
    \begin{equation}
        V_\mathrm{eff}(x) = U_F[\Delta(x)]-\frac{i\hbar\Gamma}{2}\cdot\frac{\Omega^2}{\Gamma^2+4\Delta(x)^2}
    \end{equation}
\end{itemize}

Simulations are executed across a single lattice period using periodic boundary conditions, and using a spatial discretization step of $\delta x = \lambda / 1000$, and a temporal sampling interval of $\delta t = \Gamma^{-1}$. Final observables are obtained by averaging over $10^4$ independent Monte Carlo trajectories for each parameter configuration $(s_\mathrm{780}, \Delta_\mathrm{780})$.

\subsection{Calibration of the excited state modulation amplitude}\label{app:calibration}

The excited state energy is modulated by an optical lattice which addresses the transition 5P$_{3/2}-$4D$5/2$, leading to the energy shift $U_e(x)=V_e \cos^2(kx)$, and the local detuning $\Delta(x)=\Delta_\mathrm{780}-U_e(x)/\hbar$. The calibration of the amplitude $V_e$ is realized by driving the $g-e$ transition with short pulses of scanned $\Delta_\mathrm{780}$. As the $g-e$ transition being a repumping transition, the number of repumped atoms reveals the properties of the modulation $U_e(x)$ (Fig. \ref{fig:bat_scheme}), through the local scattering rate $R(\Delta_\mathrm{780},x)$:

\begin{equation}\label{eq:localSponRate}
    R(\Delta_\mathrm{780},x) = \Gamma\rho_{ee} = \frac{\Gamma}{2}\frac{s_\mathrm{780}}{1+4\left[\Delta(x)/\Gamma\right]^2+s_\mathrm{780}}
\end{equation}

and the total repumping rate can be approximated by:

\begin{equation}\label{eq:countRep_exact}
    \frac{dN}{dt} \propto \int_{0}^{L} n(x)R(\Delta_\mathrm{780},x)dx
\end{equation}
where $L$ denotes the longitudinal size of the cloud. For $L$ much larger than the lattice constant $l$, we can further approximate the longitudinal density distribution $n(x,t)$ with its average $\bar{n}$, simplifying the expression as:

\begin{equation}\label{eq:N_R}
    \frac{dN}{dt} \propto \frac{\bar{n}L}{l}\int_{0}^{l} R(\Delta_\mathrm{780},x)dx \propto \int_{0}^{l} R(\Delta_\mathrm{780},x)dx
\end{equation}
which turns out to be a functional of $R(\Delta_\mathrm{780},x)$, and is determined by the choice of $\Delta_\mathrm{780}$. For infinitely short repumper pulses, the measured atom number $N(\Delta_\mathrm{780})$ should be proportional to $dN/dt$. Experimentally, we observe that the effect of finite pulse duration can be taken into account by introducing an effective linewidth broadening ($\Gamma\rightarrow b\Gamma$) to $R(\Delta_\mathrm{780},x)$ \cite{huang2026}:

\begin{equation}\label{eq:localSponRate_broad}
    R_b(\Delta_\mathrm{780},x) = \frac{\Gamma}{2}\frac{s_\mathrm{780}}{1+4\left[\Delta(x)/b\Gamma\right]^2+s_\mathrm{780}}
\end{equation}
With the effective correction, we measured the repumped atom number $N(\Delta_\mathrm{780})$ that is proportional to the functional $\int R_b(x)dx$, as shown in Fig. \ref{fig:bat_scheme}b. The curvature of the sinusoidal leads to a bat-like curve, and the expansion between the two wings calibrates the amplitude $V_e\approx 20\Gamma$.

\bibliography{biblio}

\end{document}